\documentclass[longauth]{aa}  

\usepackage{graphicx}
\usepackage{txfonts}
\usepackage{aas_macros}

\usepackage{natbib}
\bibpunct{(}{)}{;}{a}{}{,}

\usepackage{lipsum}
\usepackage{enumerate}
\usepackage{multirow}
\usepackage{subcaption}         
\usepackage{lscape}             
\usepackage{placeins}           

\usepackage[colorlinks=true, citecolor=blue]{hyperref}

\newcommand{\teff}{\ensuremath{T_{\rm eff}}}
\newcommand{\teq}{\ensuremath{T_{\rm eq}}}

\newcommand{\juliet}{\texttt{juliet}}
\newcommand{\zaspe}{\texttt{ZASPE}}
\newcommand{\ceres}{\texttt{CERES}}

\newcommand{\mjup}{\ensuremath{{\rm M_{J}}}}
\newcommand{\mearth}{\ensuremath{{\rm M}_{\oplus}}}
\newcommand{\mplanet}{\ensuremath{{\rm M_P}}}
\newcommand{\rjup}{\ensuremath{{\rm R_J}}}

\newcommand{\rstar}{\ensuremath{{\rm R}_{\star}}}
\newcommand{\mstar}{\ensuremath{{\rm M}_{\star}}}
\newcommand{\lstar}{\ensuremath{{\rm L}_{\star}}}

  \newcommand{\mstA}   {\ensuremath{1.386_{-0.071}^{+0.071}}}
  \newcommand{\rstA}   {\ensuremath{1.856_{-0.079}^{+0.079}}}
  \newcommand{\rhostA} {\ensuremath{306_{-39}^{+46}}}
  \newcommand{\lstA}   {\ensuremath{4.7_{-0.2}^{+0.2}}}
  \newcommand{\ageA}   {\ensuremath{2.8_{-0.6}^{+0.6}}}
  \newcommand{\AvA}    {\ensuremath{0.08_{-0.05}^{+0.05}}}
  \newcommand{\stname} {TIC65910228}
  
  \newcommand{\teffA}  {\ensuremath{6235.0_{-175}^{+175}}}
  \newcommand{\fehA}   {\ensuremath{0.14_{-0.05}^{+0.05}}}
  \newcommand{\loggA}  {\ensuremath{4.043_{-0.02}^{+0.02}}}
  \newcommand{\vsiniA} {\ensuremath{8.7_{-0.5}^{+0.5}}}

  \newcommand{\plname}{TIC65910228b}
  \newcommand{\aA}     {\ensuremath{0.697_{-0.012}^{+0.011}}}
  \newcommand{\mplA}   {\ensuremath{4.554_{-0.247}^{+0.255}}}
  \newcommand{\rplA}   {\ensuremath{1.088_{-0.057}^{+0.061}}}
  \newcommand{\rhoplA} {\ensuremath{4.374_{-0.674}^{+0.800}}}
  \newcommand{\teqA}   {\ensuremath{582_{-8}^{+8}}}
  \newcommand{\rprs}   {\ensuremath{0.0602_{0.0019}^{0.0022}}}
  \newcommand{\plbA}   {\ensuremath{0.8135_{0.0260}^{0.0200}}}
  \newcommand{\plaR}   {\ensuremath{80.812_{0.698}^{0.680}}}
  \newcommand{\plinc}   {\ensuremath{89.53_{0.01}^{0.01}}}
  \newcommand{\tdur}   {\ensuremath{11.61_{0.41}^{0.47}}}
  \newcommand{\gplA}   {\ensuremath{95.1_{10.4}^{11.8}}}

  \newcommand{\plecc}   {\ensuremath{0.25_{-0.04}^{+0.03}}}
  \newcommand{\plperA} {\ensuremath{180.524140_{-0.001048}^{+0.000978}}}
  \newcommand{\rhopost}   {\ensuremath{306.34_{-7.87}^{+7.81}}}
  \newcommand{\plephem}   {\ensuremath{2459209.229716_{-0.002220}^{+0.002145}}}
  \newcommand{\plruno}   {\ensuremath{0.8757_{-0.0173}^{+0.0133}}}
  \newcommand{\plrdos}   {\ensuremath{0.0602_{-0.0022}^{+0.0019}}}
  \newcommand{\plK}   {\ensuremath{131.833_{-5.866}^{+5.653}}}
  \newcommand{\plomega}   {\ensuremath{279.13_{-2.63}^{+3.31}}}

  \newcommand{\mfluxtess}   {\ensuremath{196_{327}^{+317} \times 10^{-7}}}
  \newcommand{\sigmatess}   {\ensuremath{2_{-2}^{+13}}}
  \newcommand{\qunotess}   {\ensuremath{0.519_{-0.165}^{+0.248}}}
  \newcommand{\qdostess}   {\ensuremath{0.406_{-0.274}^{+0.353}}}
  \newcommand{\gpsigmatess}   {\ensuremath{100_{-19}^{+29} \times 10^{-6}}}
  \newcommand{\gprhotess}   {\ensuremath{1.209_{-0.335}^{+0.503}}}

  \newcommand{\mfluxhatpi}   {\ensuremath{46_{223}^{+233} \times 10^{-7}}}
  \newcommand{\sigmahatpi}   {\ensuremath{0_{-0}^{+12}}}
  \newcommand{\qunohatpi}   {\ensuremath{0.276_{-0.200}^{+0.337}}}
  \newcommand{\qdoshatpi}   {\ensuremath{0.431_{-0.308}^{+0.352}}}
  \newcommand{\diluthatpi}   {\ensuremath{0.904_{-0.066}^{+0.062}}}

  \newcommand{\mfluxomes}   {\ensuremath{2327_{1663}^{+1671} \times 10^{-7}}}
  \newcommand{\sigmaomes}   {\ensuremath{1_{-1}^{+180}}}
  \newcommand{\qunoomes}   {\ensuremath{0.574_{-0.279}^{+0.265}}}
  \newcommand{\qdosomes}   {\ensuremath{0.406_{-0.291}^{+0.315}}}

  \newcommand{\muferos}   {\ensuremath{20027_{8}^{+7}}}
  \newcommand{\sigmaferos}   {\ensuremath{7_{-6}^{+17}}}
  \newcommand{\mupspec}   {\ensuremath{20063_{-4}^{+4}}}
  \newcommand{\sigmapspec}   {\ensuremath{19_{-4}^{+4}}}

\begin{document}

   \title{TIC65910228b: A single-transit discovery of a massive long-period warm Jupiter with TESS}



   \author{Felipe I. Rojas\inst{1}
        \and Rafael Brahm\inst{2,3}
        \and Mat\'ias I. Jones \inst{4}
        \and M\'arcio Catelan \inst{1,3}
        \and Jozef Liptak \inst{7,8}
        \and Lorena Acuña \inst{5}
        \and Jan Eberhardt \inst{5}
        \and N\'estor Espinoza \inst{6}
        \and Thomas Henning \inst{5}
        \and Andr\'es Jord\'an \inst{2,3}
        \and Yared Reinarz \inst{5}
        \and Marcelo Tala Pinto \inst{9}
        \and Trifon Trifonov \inst{5,10,11}
        \and Michaela V\'{i}tkov\'{a} \inst{7,12}
        \and Luca Antonucci \inst{13,14}
        \and Gaspar Bakos \inst{15}
        \and Attila B\'{o}di \inst{15}
        \and Gavin Boyle \inst{16,17}
        \and Zolt\'{a}n Csubry \inst{15}
        \and Joel Hartman \inst{15}
        \and Jan Janík \inst{12}
        \and Petr Kab\'{a}th \inst{7}
        \and Anthony Keyes \inst{15}
        \and Markus Roth \inst{18}
        \and Petr Škoda \inst{7}
        \and Alton Spencer \inst{19}
        \and Vincent Suc \inst{2,3,16}
        \and Geert Jan Talens \inst{20}
        \and Jan Vaclavik \inst{21}
        \and Leonardo Vanzi \inst{13,14}
        }

   \institute{Instituto de Astrof\'isica, Pontificia Universidad Cat\'olica de Chile, Av. Vicu\~na Mackenna 4860, 7820436 Macul, Santiago, Chile\\
             \email{firojas@uc.cl}
            \and Facultad de Ingenier\'ia y Ciencias, Universidad Adolfo Ib\'{a}\~{n}ez, Av. Diagonal las Torres 2640, Pe\~{n}alol\'{e}n, Santiago, Chile
            \and Millennium Institute for Astrophysics, Nuncio Monse\~{n}or Sotero Sanz 100, Of. 104, Providencia, Santiago, Chile
            \and European Southern Observatory (ESO), Alonso de C\'ordova 3107, Vitacura, Casilla 19001, Santiago, Chile
            \and Max-Planck-Institut für Astronomie, Königstuhl 17, D-69117 Heidelberg, Germany
            \and Space Telescope Science Institute, 3700 San Martin Drive, Baltimore, MD 21218, USA
            \and Astronomical Institute of the Czech Academy of Sciences, Fri\v{c}ova 298, CZ-25165 Ond\v{r}ejov, Czech Republic
            \and Astronomical Institute of Charles University, V Hole\v{s}ovi\v{c}k\'{a}ch 2, CZ-180 00 Prague, Czech Republic
            \and Department of Astronomy, McPherson Laboratory, The Ohio State University, 140 W 18th Ave, Columbus, Ohio 43210, USA
            \and Department of Astronomy, Sofia University ``St Kliment Ohridski'', 5 James Bourchier Blvd, BG-1164 Sofia, Bulgaria
            \and Landessternwarte, Zentrum f\"ur Astronomie der Universit\"at Heidelberg, K\"onigstuhl 12, D-69117 Heidelberg, Germany
            \and Department of Theoretical Physics and Astrophysics, Faculty of Science, Masaryk University, Kotl\'{a}\v{r}sk\'{a} 2, CZ-61137 Brno, Czech Republic
            \and Center of Astro Engineering, Pontificia Universidad Católica de Chile, Av. Vicuña Mackenna 4860, 782-043 Santiago, Chile
            \and Department of Electrical Engineering, Pontificia Universidad Católica de Chile, Av. Vicuña Mackenna 4860, 782-043 Santiago, Chile
            \and Department of Astrophysical Sciences, Princeton University, 4 Ivy Lane, Princeton, NJ 08544, USA
            \and El Sauce Observatory --- Obstech, Coquimbo, Chile
            \and Cavendish Laboratory, J. J. Thomson Avenue, Cambridge, CB3 0HE, UK
            \and Thüringer Landessternwarte, D-07778 Tautenburg, Germany
            \and Western Connecticut State University, Danbury, CT 06810, USA
            \and Denys Wilkinson Building, Department of Physics, University of Oxford, OX1 3RH, UK
            \and Institute of Plasma Physics of the CAS, TOPTEC department, U Slovanky 2525/1a 182 00 Prague 8, Libeň, Czech Republic
             }
   \date{Received January 31, 2026}

 
  \abstract
   {Warm Jupiters are excellent case studies for the investigation of giant planet internal structures and formation theories. However, the sample of long-period transiting giants is still small today for a better understanding of this population.}
   {Starting from a single transit found in the Transiting Exoplanet Survey Satellite (TESS) data, we confirm the planetary nature of the signal and measure its orbital parameters, mass, and radius. We put this system in the context of long-period giant transiting planets and analyzed the viability to sustain atmospheric or dynamical follow-up.}
   {We carried out a spectroscopic follow-up using FEROS and PLATOSpec to obtain precise radial velocities. We added a photometric follow-up with HATPI and Observatoire Moana to obtain a more precise estimate of the orbital period. We derived the orbital and physical parameters through a joint analysis of this data.}
   {We report the discovery and characterization of \plname , a transiting warm Jupiter with a mass of $4.554\pm0.255$ \mjup and a radius of $1.088\pm0.061$ \rjup, orbiting an evolved F-type star every $\sim180.52$ days in an eccentric orbit ($e=0.25\pm 0.04$).}
   {This planet joins a still under-explored population of long-period ($P>100$) massive ($M_p > 4$ \mjup) transiting giant planets, being one of the few with a mild eccentricity. This target is a nice example of the potential of single-transit events to populate this region of the parameter space.}


   \maketitle
   \nolinenumbers

\section{Introduction}
In recent decades, thousands of exoplanets have been discovered. Among these, warm Jupiters (WJ; radii greater than four times Earth's and orbital periods longer than 10 days) represent a valuable sample for studying the formation and evolution of planetary systems. Unlike hot Jupiters \citep[HJ;][]{mayorqueloz95}, warm Jupiters are located far enough from their host stars to avoid inflation caused by stellar irradiation, which complicates internal structure studies \citep[e.g.][]{thorngren2016, fortney2021, acuna2024}, as the proposed mechanisms are not well understood \citep{sarkis2021}. This distance also prevents tidal circularization, a process that can alter their original orbital parameters, such as eccentricity and obliquity, which provide insights into their formation and migration scenarios \citep{albrecht2022}. In this context, different mechanisms have been proposed that could explain the observed population of warm Jupiters \citep[e.g.][]{gold1980, rf1996, insitu, kozai, lidov, secular}. Their mass, size, and orbital period make them ideal targets for combining transit and radial velocity observations around bright stars.

Since its launch in 2018, the Transiting Exoplanet Survey Satellite \citep[TESS;][]{tess} has been surveying the entire sky in search of transiting exoplanets, focusing on regions of 24 by 96 degrees every 27 days.
However, as the orbital period increases, the likelihood of observing a transit decreases, posing a significant challenge for characterizing systems with long orbital periods. Given this strategy, most warm Jupiters in TESS are expected to be observed as single transiting events or with interrupted observations, preventing a direct measurement of their orbital periods. Simulations show that 76\% of warm Jupiters with periods longer than 100 days should be observed as single transiters in years $1-3$ \citep{rodel2024}. In the first two years of data, it is expected that between 500 and 1000 single-transit events will be observed \citep{cooke, villanueva}.

To overcome this challenge, the Warm gIaNts with TESS (WINE) collaboration has been aiming to identify and characterize such planets through ground-based follow-up observations of TESS candidates \citep{brahm2023,jones2024,eberhardt2025}. In particular, for single transiters, spectroscopic monitoring is crucial to constrain the orbital period if no additional transits are found and to rule out false-positive scenarios. 

In this study, we present the discovery and detailed characterization of \plname , a warm Jupiter orbiting an evolved F-type star with an orbital period of 180.5 days. The planet was first detected as a single-transit signal in TESS data and later confirmed through high-precision radial velocity measurements from FEROS and PLATOSpec, along with follow-up ground-based transit observations conducted with HATPI and at Observatoire Moana. This system constitutes an important addition to the still limited sample of well-characterized warm Jupiters with accurately determined masses and radii.

The paper is structured as follows: In Section \ref{sec:obs}, we present the photometric and spectroscopic data that were collected. Section \ref{sec:analysis} outlines the analysis conducted on this data. The results and discussion are provided in Section \ref{sec:results}, and we conclude with a summary in Section \ref{sec:conclusions}.

\section{Observations}
\label{sec:obs}

Besides the data from TESS, a photometric follow-up from the ground is needed to i) have a more precise constraint on the period and ii) observe the transit with better resolution to confirm the origin of the signal and discard blended scenarios. A spectroscopic follow-up was also performed to measure precise radial velocities and characterize the stellar host.

A summary of the photometric and spectroscopic observations used in the analysis can be found in Table \ref{tab:followup}.

\subsection{Photometry}

\begin{table}
    \caption{Summary of ground-based follow-up observations.}
    \label{tab:followup}
    \centering 
    \begin{tabular}{lccc}
    \hline\hline
    Telescope & Filter & Coverage & Date \\
    \hline
    TESS & \multirow{2}{*}{TESS} & \multirow{2}{*}{Full} & 2018/12/18 \\
    Sector 33 & & & to 2019/01/13 \\[0.1cm]
    \multirow{2}{*}{HATPI} & Asahi Spectra & \multirow{2}{*}{Egress} & \multirow{2}{*}{2024/12/09} \\
    &  430--890 nm & & \\[0.1cm]
    OM-ES1 & r & Egress & 2025/12/05 \\ \hline\vspace{1ex}
    \end{tabular}
    \begin{tabular}{lcc}
    \hline\hline
    Instrument & Observations & Date \\
    \hline
    FEROS & 6 & 2024/12/10 --- 2025/03/30 \\[0.1cm]
    PLATOSpec & 42 & 2024/12/13 --- 2026/01/05 \\ \hline
    \end{tabular}
\end{table}

\subsubsection{TESS}
\stname\ was observed during the first year of TESS Mission \citep{tess} in Sector 7, with a cadence of 30 minutes. During the first extended mission, observations were conducted in Sectors 33 and 34 at a 10-minute cadence. In the second extended mission, it was observed in Sector 61 with a cadence of 2 minutes. In year 7, it was observed in Sectors 87 and 88.

A single transit was identified in Sector 33 and reported as a CTOI in the Exoplanet Follow-up Observation Program (ExoFOP; DOI: 10.26134/ExoFOP5). TESS-SPOC \cite{SPOC} light curve from Sector 33 was employed for this analysis, in particular using the \texttt{PDCSAP} flux.

\begin{figure}
    \centering
    \includegraphics[width=\columnwidth]{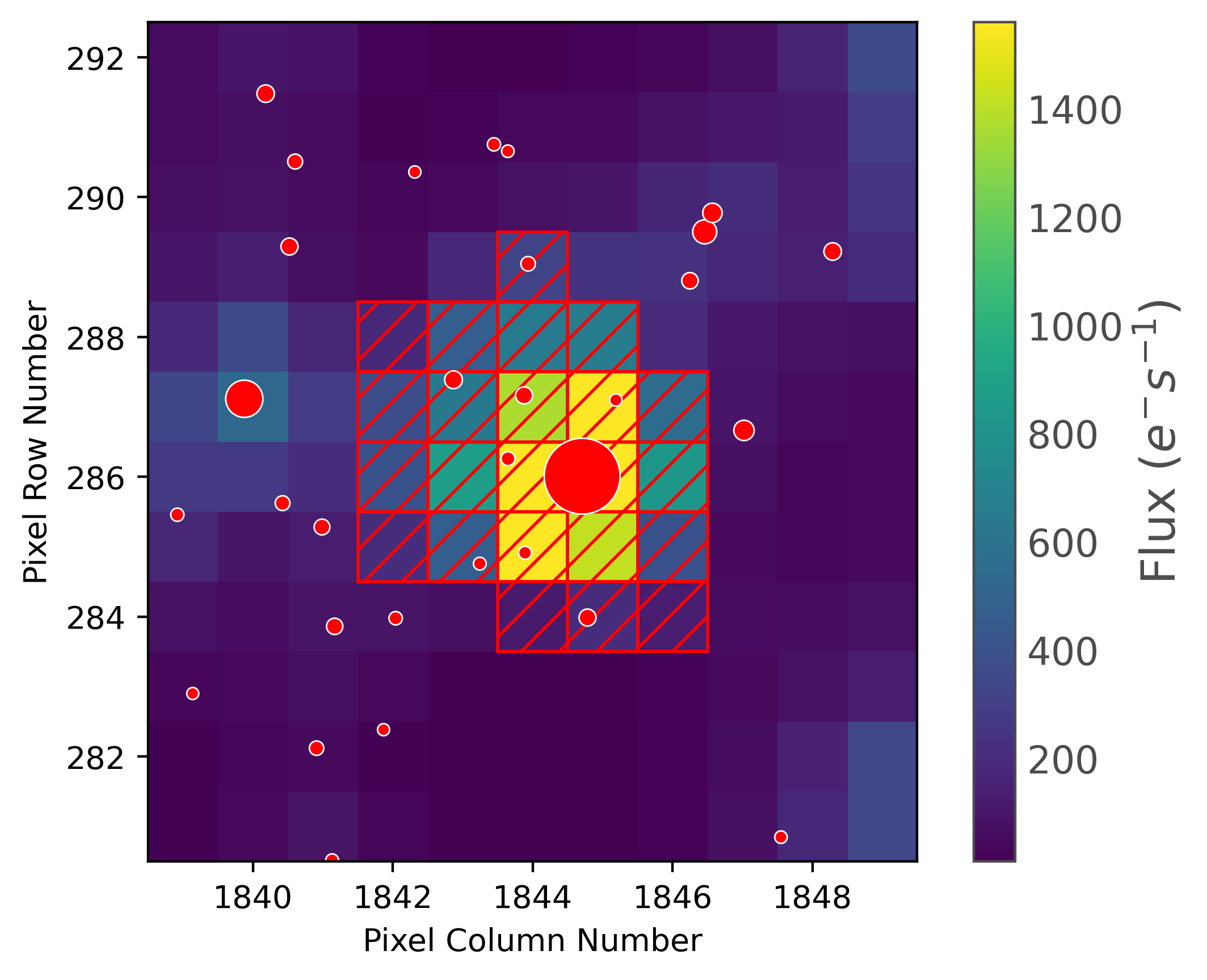}
    \caption{TESS Target Pixel File preview of \stname . Aperture used in TESS-SPOC light curve corresponds to the overplotted red dashed area. Gaia DR3 sources up to 7 magnitudes of difference are overplotted in red dots. Dot size is proportional to brightness. }
    \label{fig:tessgaia}
\end{figure}

Figure \ref{fig:tessgaia} shows the TESS field of \stname\ in Sector 33 with Gaia \citep{gaia3} sources overplotted and the aperture used in the TESS-SPOC light curve. The light curve is plotted in Figure \ref{fig:LCs}.

\subsubsection{HATPI}
HATPI is a wide-field photometric instrument located at Las Campanas Observatory (LCO). The HATPI system comprises a mosaic of 64 identical camera–lens units mounted together on a single platform, enabling continuous imaging of the entire sky visible from LCO with cadences of 30 and 45 seconds and a pixel scale of 20 arcsec pix$^{-1}$ (Bakos et al., 2025, in preparation).

High-precision light curves were generated using aperture and image subtraction photometry. This data is available in three formats: raw photometry, detrended via External Parameter Decorrelation (EPD) and with systematic correction using Trend Filtering Algorithm \citep[TFA;][]{tfa} in the EPD light curve. EPD removes variability that correlates with instrumental and non-astrophysical variables, in this case the image position of the source, the horizontal angle, the airmass and the shape of the point-spread function. We used the subtracted photometry TFA light curve for this work.

The target was observed in three different seasons, with light curves that spans from 2022/08/17 to 2023/06/02, 2023/08/17 to 2024/06/02, and 2024/08/17 to 2025/06/02. An egress of a \plname\ transit was observed by HATPI on December 9, 2024, which is consistent with the period found in Section \ref{sec:gls}.

\subsubsection{Observatoire Moana}
Another egress from \plname\ was observed on December 5, 2025, with a robotic 60 cm telescope located at El Sauce station of Observatoire Moana (OM-ES). The telescope is equipped with a $2048\times2048$ pixel detector, providing a pixel scale of 0.7 arcsec pix$^{-1}$. Observations were carried out in the $r^{\prime}$ band, with exposures every 50 seconds.

The reduction process was carried out using an automated pipeline that integrates all available sources within the field. This generates light curves and artificial comparison stars for all qualifying sources, accounting for variability and magnitude differences. Additionally, it uses an optimization algorithm that accounts for the spatial separation between the targets and other sources.

This egress was predicted using previous analysis with HATPI data and radial velocities, which allowed us to constrain the period and refine the ephemeris to schedule the observations.

\subsection{Radial Velocities}

\subsubsection{FEROS}
Six observations were conducted with the FEROS spectrograph \citep{feros} mounted at the 2.2m MPG telescope in La Silla Observatory. The adopted exposure time was 900 seconds. Data were observed between December 10, 2024, and March 30, 2025.

Data were processed using the \ceres\ pipeline \citep{ceres}, which calculates the cross-correlation function (CCF), obtaining precision radial velocities, bisector span, and full width at half maximum (FWHM) of the CCF. The spectra obtained had a mean signal-to-noise ratio of 84.

\subsubsection{PLATOSpec}
We observed \stname\ 42 times with PLATOSpec \citep{platospec} between December 13, 2024, and January 5, 2026. PLATOSpec is a high-resolution fiber-fed echelle spectrograph designed to support space missions such as PLATO and Ariel. It is installed in the 1.52m telescope at La Silla Observatory, covering a spectral range of 380 to 680 nm with a resolving power of $R=70000$. The exposures lasted 900 seconds, yielding a mean signal-to-noise ratio of 53.

Observations were reduced using the \ceres\ pipeline, measuring radial velocities, bisector span, and FWHM.

\section{Analysis}
\label{sec:analysis}
\subsection{Stellar Parameters}
For the spectroscopic properties, we followed the same procedure in \citep{brahm2019b}. The first step consisted of analyzing the co-added spectra to derive $T_{\mathrm{eff}}$, $\log g$, [Fe/H], and $v\sin i$. This is done using the \zaspe\ package \cite{zaspe}, which determines these parameters by comparing the data with a grid of synthetic spectral models.

After this, we performed a spectral energy distribution (SED) fit using available broad-band photometry, in combination with \texttt{parsec} stellar evolutionary models \citep{parsec} and parallax from Gaia DR3 \citep{gaiadr3}. In this stage, the previously derived $T_{\mathrm{eff}}$ was used as a prior, while the metallicity is fixed. In this way, we obtain an improved estimate of the surface gravity, which is then fixed for another iteration of \zaspe\ .

These two steps were repeated until convergence was reached. From this analysis, we found that \stname\ is an F-type metal-rich evolved star.

Table \ref{tab:stellarparams} resumes all stellar parameters found in the literature and derived from the high-resolution spectra. These results were adjusted according to the guidelines in \cite{tayar2022}, where uncertainties of \teff, \lstar , \rstar , \mstar\, and age have been raised by 2.4\%, 2\%, 4.2\%, 5\%, and 20\% in quadrature, respectively.

\begin{table*}
\caption{Stellar parameters of \stname.}
\label{tab:stellarparams}
\centering
  \begin{tabular}{lcr}
   \hline
   \hline
     Parameter &  &  Source\\
   \hline
Identifying Information & & \\
~~~TIC ID & \stname\ &   TIC$^a$\\
~~~GAIA ID & 5606317297918628992 &\textit{Gaia} EDR3$^b$ \\
~~~2MASS ID & J07145152-2925498 & 2MASS$^c$\\
~~~R.A. (J2015.5, h:m:s) & $7^h14^m51.52^s$ & \textit{Gaia} EDR3$^b$ \\
~~~DEC (J2015.5, d:m:s) & $-29^\circ25'49.91''$ & \textit{Gaia} EDR3$^b$\\
Proper motion and parallax & & \\
~~~$\mu_\alpha \cos \delta$ (mas yr$^{-1}$) & -13.566 $\pm$ 0.011 & \textit{Gaia} EDR3$^b$ \\
~~~$\mu_\delta$ (mas yr$^{-1}$) & -21.083 $\pm$ 0.013 & \textit{Gaia} EDR3$^b$ \\
~~~Parallax (mas) & 3.74 $\pm$ 0.013 & \textit{Gaia} EDR3$^b$ \\
Spectroscopic properties  & & \\
~~~$T_\textnormal{eff}$ (K) & \teffA & ZASPE$^d$\\
~~~Spectral Type & F & ZASPE$^d$\\
~~~[Fe/H] (dex) & \fehA & ZASPE$^d$\\
~~~$\log g_*$ (cgs)& \loggA & ZASPE$^d$\\
~~~$v\sin(i_*)$ (km/s)& \vsiniA & ZASPE$^d$\\
Photometric properties  & & \\
~~~$T$ (mag)& $9.734 \pm 0.004$ & TIC$^a$\\
~~~$G$ (mag)& $10.110 \pm 0.0003$ & \textit{Gaia} EDR3$^b$ \\
~~~$B$ (mag)& $10.819\pm 0.064$ & Tycho-2$^e$\\
~~~$V$ (mag)& $10.225\pm 0.004$  &Tycho-2$^e$\\
~~~$J$ (mag)& $9.251\pm 0.026$ & 2MASS$^c$\\
~~~$H$ (mag)& $9.01\pm 0.026$ & 2MASS$^c$\\
~~~$Ks$ (mag)& $8.938 \pm0.023$ & 2MASS$^c$\\
Derived properties  & & \\
\vspace{0.1cm}
~~~$M_*$ ($M_\odot$)& \mstA & PARSEC$^{*}$\\
~~~$R_*$ ($R_\odot$)& \rstA & PARSEC$^{*}$\\
~~~$L_*$ ($L_\odot$)& \lstA & PARSEC$^{*}$\\
~~~$A_v$ (mag) & \AvA & PARSEC$^{*}$\\
~~~Age (Gyr)& \ageA & PARSEC$^{*}$\\
~~~$\rho_*$ (g cm$^{-3}$)& \rhostA & PARSEC$^{*}$\\
   \hline
   \end{tabular}

\tablefoot{Logarithms given in base 10.\\
\tablefoottext{a}{\textit{TESS} Input Catalog \citep{TIC8}}; \tablefoottext{b}{\textit{Gaia} Early Data Release 3 \citep{gaia3}};\tablefoottext{c}{Two-micron All Sky Survey \citep{2MASS}};\tablefoottext{d}{Zonal Atmospheric Stellar Parameters Estimator \citep{brahm2019b,zaspe}};\tablefoottext{e}{Tycho-2 Catalog \citep{tycho2}}\\
*: PARSEC isochrones \citep{parsec}; using stellar parameters obtained from ZASPE.}
\end{table*}

\subsection{Periodical signals and stellar activity}
\label{sec:gls}
We calculated the generalized Lomb–Scargle periodogram \citep{GLS} for the PLATOSpec observations to search for periodical signals. The periodogram displays a prominent peak at approximately 180 days. Additionally, we calculated the periodogram for the bisector span and the FWHM, as these quantities serve as proxies for stellar activity. 

\begin{figure}
    \centering
    \includegraphics[width=\linewidth]{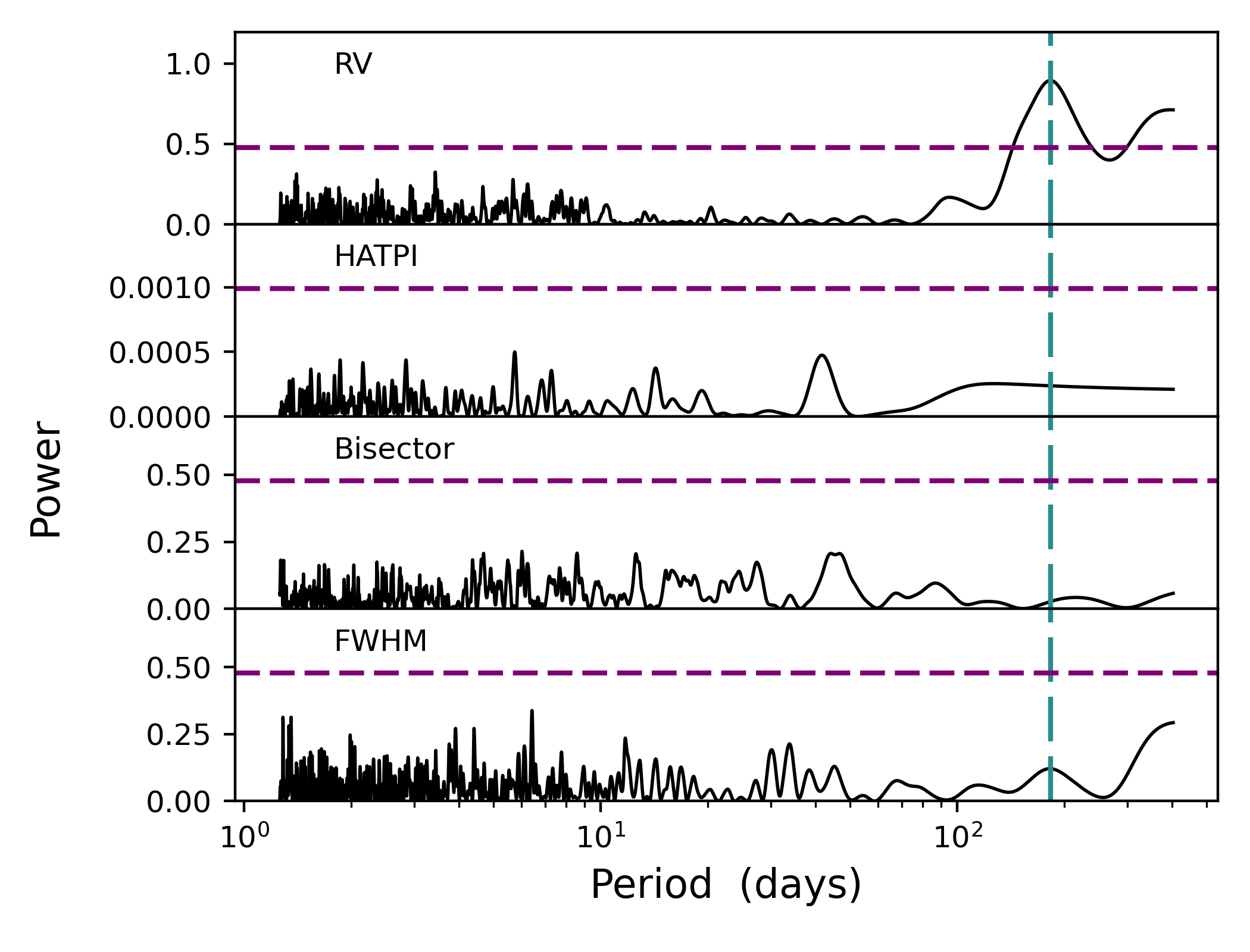}
    \caption{Periodograms displaying periodic signals in the radial velocities, HATPI photometry, radial-velocity bisector span and radial-velocity FWHM. The green vertical line indicates the optimal period derived from radial-velocity data. Horizontal dashed line corresponds to the 1\% false alarm probability.}
    \label{fig:GLS}
\end{figure}

Figure \ref{fig:GLS} shows the periodogram of radial velocities, bisector span, and FWHM derived from the PLATOSpec observations, together with the GLS periodogram of the HATPI light curve, which can uncover variability linked to stellar rotation.

\begin{figure}
    \centering
    \includegraphics[width=\linewidth]{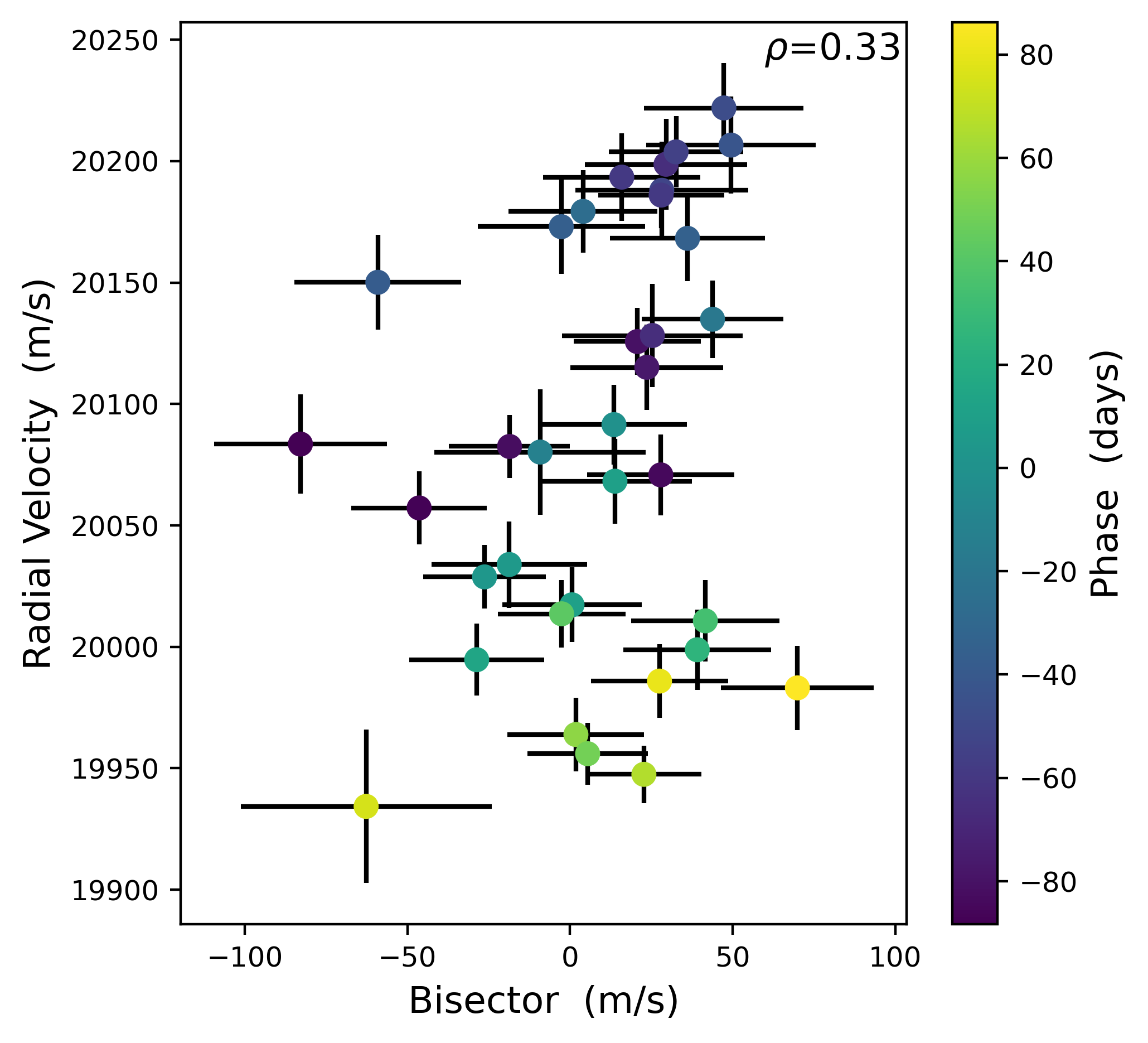}
    \caption{Bisector vs radial velocity scatter plot. Spearman's rank correlation coefficient $\rho$ shows no significant correlation between the two variables, ruling out the possibility that the RV signal has an activity origin traced by the Bisector.}
    \label{fig:BIS}
\end{figure}

The absence of notable features in the activity indicators, aligned with those detected in the radial velocity data, indicates that the observed signal is most likely of planetary origin. Also, bisectors don't show a significant correlation with radial velocities (see Figure \ref{fig:BIS}).

\subsection{Joint-fit}
We performed a joint fit of all the collected \plname\ data using \juliet. This software relies on nested sampling \citep[NS;][]{ns} and incorporates several state-of-the-art models and tools. In particular, we used \texttt{batman} \citep{batman} to model the transits, \texttt{radvel} \citep{radvel} for the radial-velocity data, and \texttt{george} \citep{george} together with \texttt{celerite} \citep{celerite} to describe the Gaussian processes (GP). Nested sampling was carried out with the \texttt{dynesty} library \citep{dynesty1} with 1000 live points.

The orbital parameters used for the model were the orbital period $P$, the ephemeris of the transit $t_0$, stellar density $\rho_{\star}$ (instead of $a/R_{\star}$), radial velocity semi-amplitude $K$, orbital eccentricity $e$, and argument of periastron passage $\omega$. We adopted the $r_1$, $r_2$ parametrization from \citep{r1r2} instead of fitting the planet-to-star radius ratio $R_p/R_{star}$ and the impact parameter $b$ directly.

In addition, instrumental parameters are required to account for the different properties of each dataset, in this case: the dilution factor $D$, relative flux offset $m_F$, limb darkening parameters, systemic radial velocity $\mu$, jitter parameters, and possible terms that account for light curve variability.

The TESS Sector 33 light curve was resampled to a 1.6-minute cadence following the method described by \cite{binning} to correct for morphological distortions caused by the integration time.

\subsubsection{Dilution}
Due to the relatively large pixel scales of TESS (21 arcsec pix$^{-1}$) and HATPI (19.7 arcsec pix$^{-1}$), a dilution factor is typically included to account for crowding effects. However, since we used the \texttt{PDCSAP\_FLUX} from the TESS–SPOC light curves, no additional correction is required, as these fluxes are already adjusted for crowding. For HATPI, we adopted a uniform prior between 0.8 and 1, as it is expected that the dilution from this instrument shouldn't exceed 10\%. In contrast, OM-ES1 photometry is primarily limited by seeing, therefore, we fixed the dilution factor to 1 in this case.

\subsubsection{Limb darkening}
Following the recommendations of \citep{ldlaw}, we adopted a quadratic law for the transit fitting of all the photometric data, adopting the parametrization proposed by \citep{kipping2013}. These parameters $q_1$ and $q_2$ are inherent to each filter and instrument, so all three photometric sets have independent ones.

\subsubsection{Jitter and variability}
Jitter terms $\sigma$ are included for each instrument to address possible underestimated error bars and unknown sources that introduce variability in both photometry and radial velocity.

\begin{figure*}
    \centering
    \includegraphics[width=\linewidth]{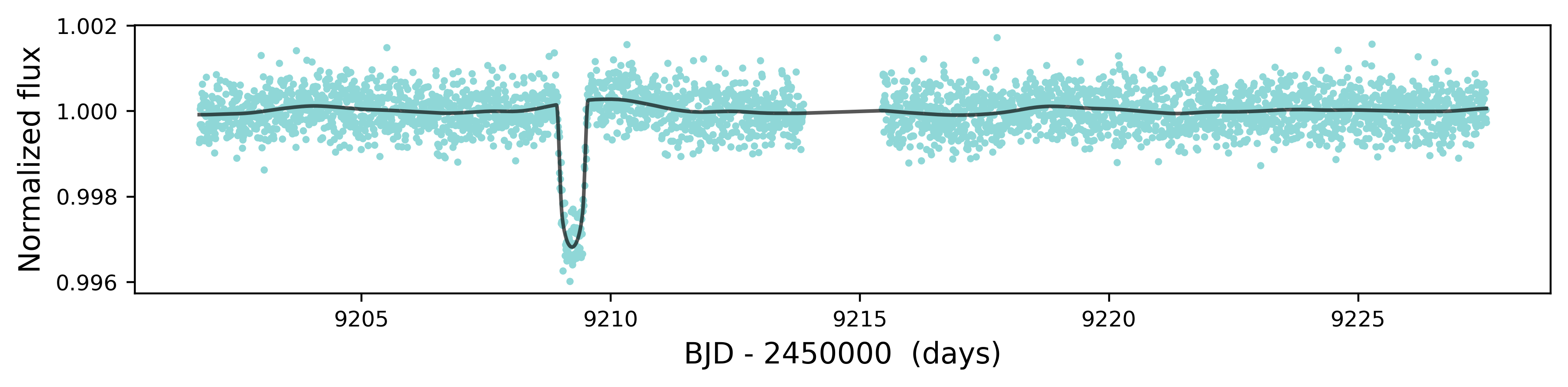}
    \caption{Light curve of TESS Sector 33. The best-fit model is overplotted in black.}
    \label{fig:TESSLC}
\end{figure*}

For photometry in particular, additional parameters were added to account for long-term trends. For TESS, a GP over time with a Matern 3/2 kernel was employed to model variations in the out-of-transit flux (see Figure \ref{fig:TESSLC}).

\begin{figure*}
    \centering
    \includegraphics[width=\linewidth]{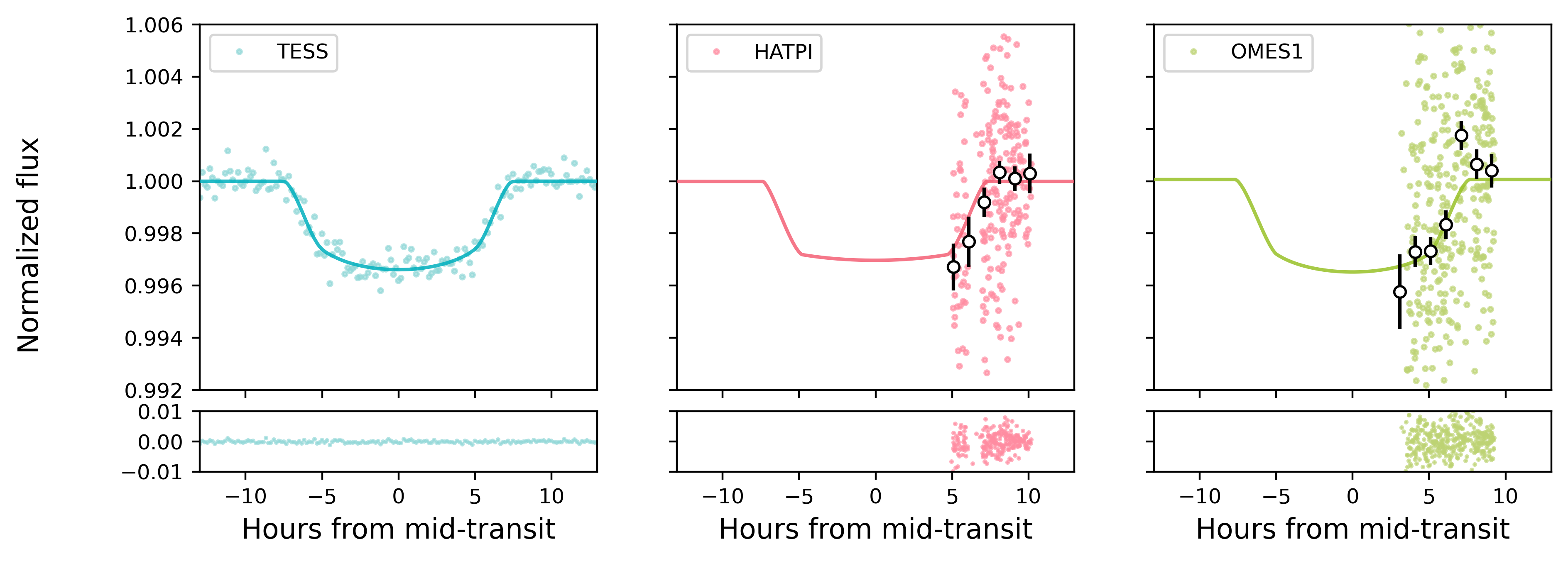}
    \caption{Top panel: Observed transits with TESS, HATPI and OM-ES1, respectively. Model from joint-fit is overplotted. White circles correspond to 1-hour bins in HATPI and OMES1 light curves. Bottom panel: Residuals after removing the transit model.}
    \label{fig:LCs}
\end{figure*}

\begin{figure*}
    \centering
    \includegraphics[width=\linewidth]{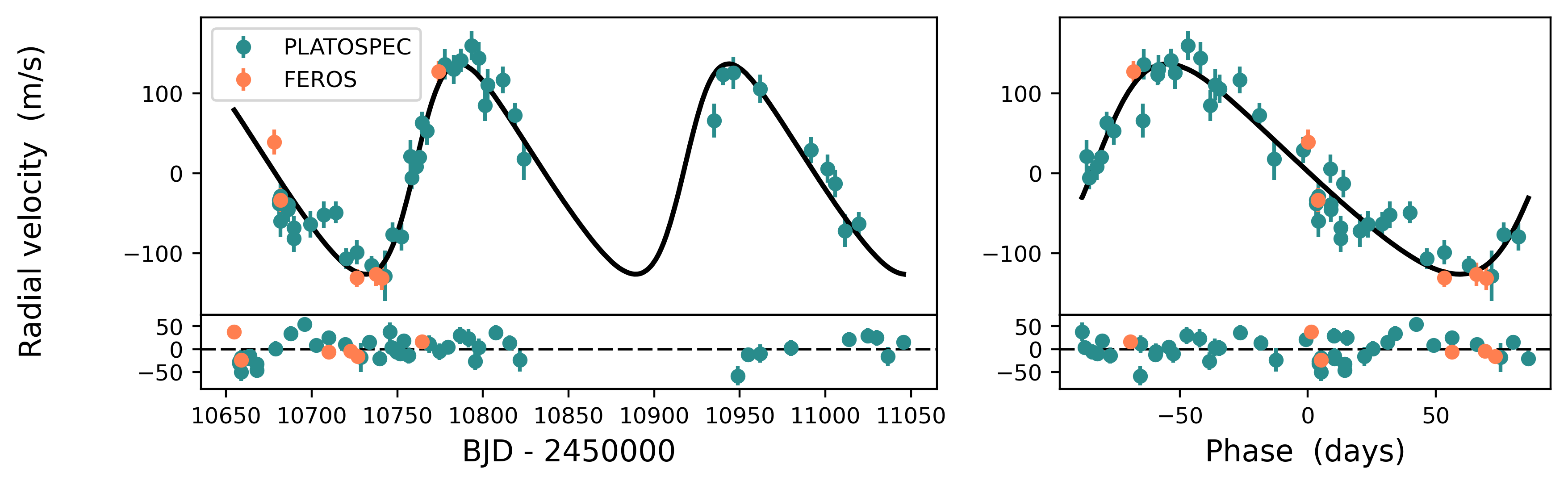}
    \caption{Left panel: Radial velocities time series obtained with FEROS and PLATOSpec. Best model is plotted in black. Residuals after removing the Keplerian model are shown in bottom. Right panel: Same but phase folded radial velocities with the inferred period from the joint-fit.}
    \label{fig:RVs}
\end{figure*}

\subsection{Model selection}
As nested sampling estimates the Bayesian evidence, it provides a procedure for comparing different models, with the one with the highest evidence being favored. In the context of this paper, we compared the following main models:

\begin{enumerate}[i]
    \item No planet: The variations in RV are explained just by noise
    \item 1 eccentric planet: The RV signal is explained by a single planet. Eccentricity is left as a free parameter.
    \item 1 eccentric planet + RV linear trend: Same as before, but adding a linear trend to the Keplerian model, coming from an external companion with a period much longer than the baseline of observations.
\end{enumerate}

\begin{figure}
    \centering
    \includegraphics[width=\columnwidth]{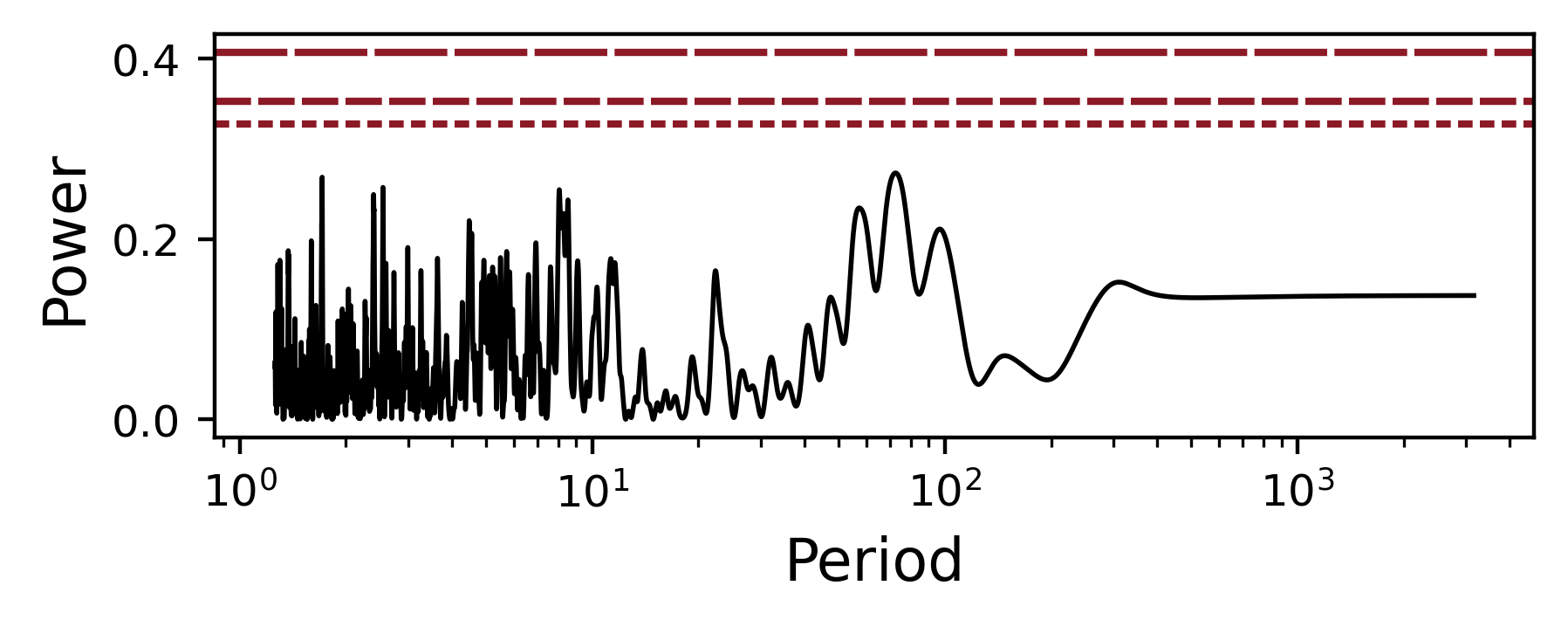}
    \caption{GLS periodogram of radial velocity residuals after removing the best fit model. Given the current data and time baseline, no significant peaks are found.}
    \label{fig:GLSres}
\end{figure}

Using Bayesian evidence, model ii is favored over model iii with a $\Delta Z = 2.2$. Although the difference is not significant, model ii is the simplest option and thus is selected. The best-fit model was subtracted from the radial velocity data, and the residuals were analyzed with a GLS periodogram, which showed no additional periodic signals (see Figure \ref{fig:GLSres}).

\begin{table*}[]
    \centering
    \begin{tabular}{lcccl}
         \hline\hline
         Parameter & Prior & Posterior & Units & Description  \\ \hline
         Star parameters & & \\ 
         ~~~$\rho_{\star}$ & $\mathcal{N}(306,8^2)$ & \rhopost & kg m$^{-3}$ & Stellar density \\[0.1cm]
         Planet parameters & & \\
         ~~~$P_b$ & $\mathcal{N}(180.5,0.5^2)$ & \plperA & days & Orbital period \\
         ~~~$t_{0,b}$ & $\mathcal{N}(2459209.23,0.5^2)$ & \plephem  & days & Time of transit center \\ 
         ~~~$r_{1,b}$ & $\mathcal{U}(0,1)$ & \plruno & & Parametrization for $p$ and $b$\\
         ~~~$r_{2,b}$ & $\mathcal{U}(0,1)$ & \plrdos & & Parametrization for $p$ and $b$\\
         ~~~$K_b$ & $\mathcal{U}(0,1000)$ & \plK & m s$^{-1}$ & Radial velocity semi-amplitude \\
         ~~~$e_b$ & $\mathcal{U}(0,0.9)$ & \plecc & & Orbital eccentricity \\
         ~~~$\omega_b$ & $\mathcal{U}(0,360)$ & \plomega & degrees & Argument of periapsis \\[0.1cm]
         Parameters for TESS & & \\ 
         ~~~$D_{\mathrm{TESS}}$ & 1 (fixed) & 1 & & Dilution factor for TESS \\
         ~~~$M_{\mathrm{TESS}}$ & $\mathcal{N}(0,0.001^2)$ & \mfluxtess & & Relative flux offset for TESS\\
         ~~~$\sigma_{\mathrm{TESS}}$ & $\mathcal{J}(10^{-1},10^4)$ & \sigmatess & ppm & Extra jitter term for TESS \\ 
         ~~~$q_{1,\mathrm{TESS}}$ & $\mathcal{U}(0,1)$ & \qunotess & & Quadratic limb-darkening parametrization \\ 
         ~~~$q_{2,\mathrm{TESS}}$ & $\mathcal{U}(0,1)$ & \qdostess & & Quadratic limb-darkening parametrization \\
         ~~~$GP_{\sigma ,TESS}$ & $\mathcal{J}(10^{-6},10^6)$ & \gpsigmatess & & Amplitude of GP component \\
         ~~~$GP_{\rho ,TESS}$ & $\mathcal{J}(10^{-6},10^6)$ & \gprhotess & & $\rho$ for GP component \\[0.1cm]
         Parameters for HATPI & & \\ 
         ~~~$D_{\mathrm{HATPI}}$ & $\mathcal{U}(0.8,1)$ & \diluthatpi & & Dilution factor for HATPI \\
         ~~~$M_{\mathrm{HATPI}}$ & $\mathcal{N}(0,0.001^2)$ & \mfluxhatpi & & Relative flux offset for HATPI\\
         ~~~$\sigma_{\mathrm{HATPI}}$ & $\mathcal{J}(10^{-1},10^4)$ & \sigmahatpi & ppm & Extra jitter term for HATPI \\ 
         ~~~$q_{1,\mathrm{HATPI}}$ & $\mathcal{U}(0,1)$ & \qunohatpi & & Quadratic limb-darkening parametrization \\ 
         ~~~$q_{2,\mathrm{HATPI}}$ & $\mathcal{U}(0,1)$ & \qdoshatpi & & Quadratic limb-darkening parametrization\\[0.1cm]
         Parameters for OM-ES1 & & \\ 
         ~~~$D_{\mathrm{OMES1}}$ & 1 (fixed) & 1 & & Dilution factor for OM-ES1 \\
         ~~~$M_{\mathrm{OMES1}}$ & $\mathcal{N}(0,0.001^2)$ & \mfluxomes & & Relative flux offset for OM-ES1\\
         ~~~$\sigma_{\mathrm{OMES1}}$ & $\mathcal{J}(10^{-1},10^4)$ & \sigmaomes & ppm & Extra jitter term for OM-ES1 \\ 
         ~~~$q_{1,\mathrm{OMES1}}$ & $\mathcal{U}(0,1)$ & \qunoomes & & Quadratic limb-darkening parametrization \\ 
         ~~~$q_{2,\mathrm{OMES1}}$ & $\mathcal{U}(0,1)$ & \qdosomes & & Quadratic limb-darkening parametrization \\[0.1cm]
         RV parameters & & \\ 
         ~~~$\mu_{FEROS}$ & $\mathcal{U}(20000,20100)$ & \muferos & m s$^{-1}$ & Systemic velocity for FEROS \\
         ~~~$\sigma_{FEROS}$ & $\mathcal{J}(10^{-1},10^2)$ & \sigmaferos & m s$^{-1}$ & Extra jitter term for FEROS \\
         ~~~$\mu_{PLATOSPEC}$ & $\mathcal{U}(20000,20100)$ & \mupspec & m s$^{-1}$ & Systemic velocity for PLATOSpec \\
         ~~~$\sigma_{PLATOSPEC}$ & $\mathcal{J}(10^{-1},10^2)$ & \sigmapspec & m s$^{-1}$ & Extra jitter term for PLATOSpec \\
         \hline
    \end{tabular}
    \caption{Priors used for the joint analysis of \plname . $\mathcal{U}$, $\mathcal{J}$ and $\mathcal{N}$ represent uniform, log-uniform and normal distributions, respectively. Posterior values obtained are listed in the adjacent column. }
    \label{tab:priors}
\end{table*}



\section{Results}
\label{sec:results}

The analysis confirms that \stname\ is an evolved F-type star hosting a warm giant planet. This planet has a mass of $M_p=\mplA$ \mjup , a radius of $R_p=\rplA$ \rjup, and a bulk density of $\rho_p=\rhoplA$ g cm$^{-3}$. The planet is in a mildly eccentric orbit with $e=\plecc$ with a period of $P=\plperA$ days.

Results from the posterior distributions of the joint-fit are summarized in Table \ref{tab:priors}. Best model fits of the transits can be seen in Figure \ref{fig:LCs} and for radial velocities in Figure \ref{fig:RVs}. All derived parameters are listed in Table \ref{tab:finalparams}.

\begin{table*}[]
    \centering
    \begin{tabular}{lcl}
         \hline\hline
         Parameter name & Posterior estimate & Description  \\ \hline
         Transit parameter & \\
         ~~~$R_p/R_{\star}$ & \rprs & Planetary radius to star radius\\
         ~~~$b$ & \plbA & Transit impact parameter \\ 
         ~~~$a/R_{\star}$ & \plaR & Normalized semi-major axis \\
         ~~~$i$ (deg) & \plinc & Orbital inclination \\
         ~~~$t_T$ (hour) & \tdur & Transit duration \\[0.2cm]
         Planet parameter & \\
         ~~~$M_p$ (\mjup ) & \mplA & Mass \\ 
         ~~~$R_p$ (\rjup ) & \rplA & Radius \\ 
         ~~~$\rho_p$ (g cm$^{-3}$ ) & \rhoplA & Bulk density \\ 
         ~~~$g_p$ (m s$^{-2}$) & \gplA & Surface gravity \\ 
         ~~~$a$ (AU) & \aA & Semi-major axis \\
         ~~~\teq (K) & \teqA & Equilibrium temperature \\ \hline
    \end{tabular}
    \caption{Transit and planetary parameters derived from the joint-fit. Median and 68\% confidence intervals from the posteriors are reported.}
    \label{tab:finalparams}
\end{table*}

\subsection{Interior modelling}
\label{sec:interior}

\begin{figure*}[h]
    \centering
    \includegraphics[width=0.65\linewidth,angle=90]{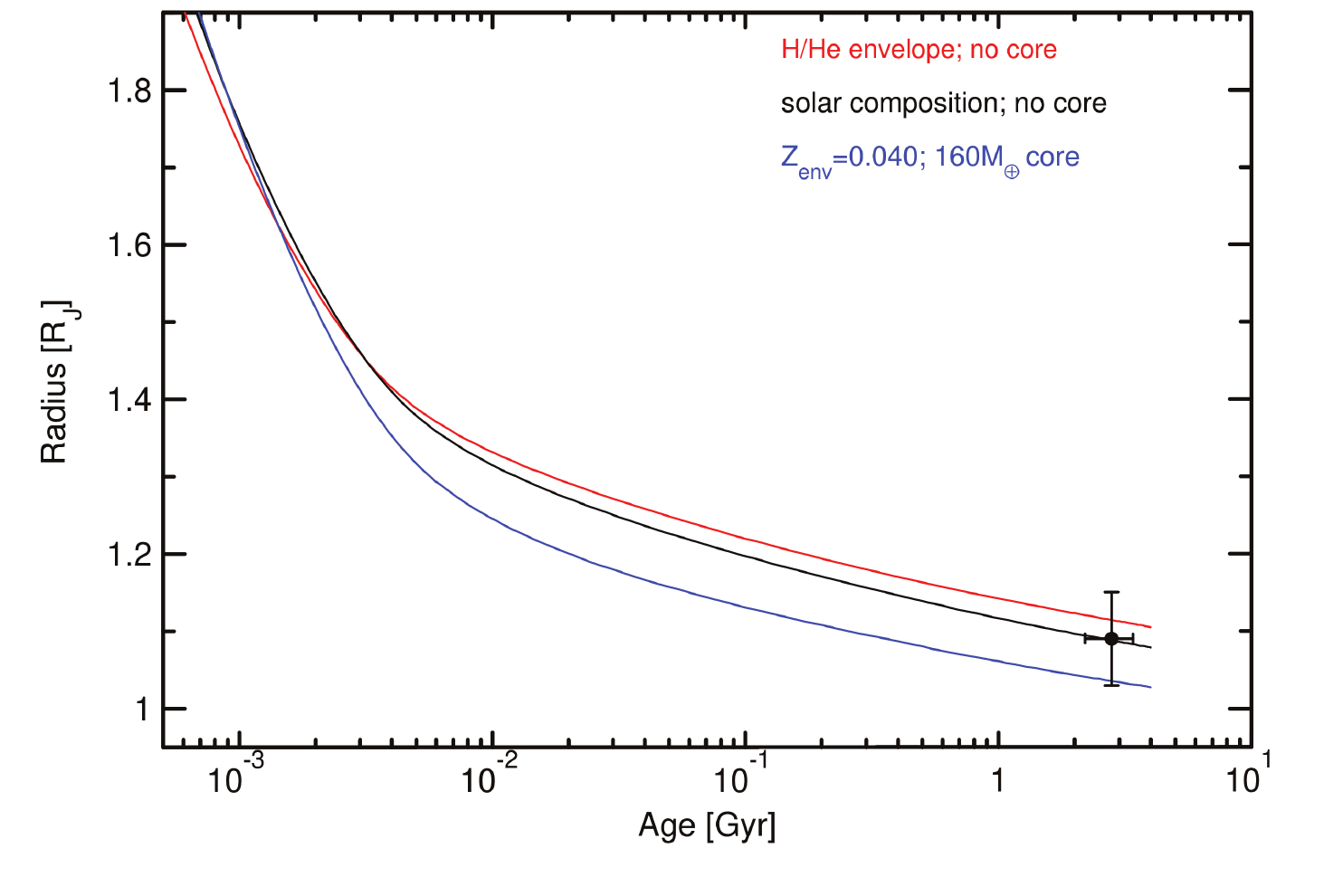}
    \caption{Position of TIC65910228\,$b$ in the age-radius diagram (black dot). Different interior models with different envelope composition and core masses over-plotted. }
    \label{fig:mesa_models}
\end{figure*}

We computed interior models of \plname\ using the Modules for Experiments in Stellar Astrophysics \citep[MESA;][]{paxton2011, paxton2013}, following the methododology presented in \cite{jones2024} and \cite{Tala2025}. 
In this case, we modeled the planet with an inert isodensity core with different masses, surrounded by a gaseous envelope with different metallicity values. 
For the core, we used a 1:1 mixture of rock and ice, with their density obtained from the \(\rho-P\) relations presented in \cite{Hubbard1989}. We also included the effect of the stellar irradiation, assuming a zero albedo for the planet's radiative atmosphere, which is updated in steps of 500 Myr.
Finally, we evolved different models, with different masses of the core, and we compared them with the position of \plname\ in the age--radius diagram. Figure \ref{fig:mesa_models} shows the position of \plname\ in the age-radius diagram, with different interior models. As can be seen, a model with solar composition for the envelope (Z = 0.015) and no core, is the one that better reproduces its position, suggesting a planet with very little heavy-elements content. We note, that the lightest possible model we could compute, that is, a pure H/He envelope with no core, does not match the 1-$\sigma$ upper radius limit. On the other hand, a model with a massive core (M$_{\rm core}$ = 160 \mearth;  $\rho_{\rm core}$ = 16 [g/cm$^3$]), matches the 1-$\sigma$ lower limit. 
These results translate into a heavy-element enrichment with respect to the host star of $Z_{p} / Z_{\star}$ = 0.7$^{+3.0}_{-0.7}$.

\subsection{\plname\ in context}
Physical parameters confirm a non-inflated radius, distinctive of a warm Jupiter. To provide context, we compare the obtained parameters with those of planets from the Exoplanet Archive\footnote{Accessed on 2026-01-21 at 08:33 UT} \citep{exoplanetarchive} with periods longer than $P=10$ days and uncertainties lower than 20\%.

\begin{figure}
    \centering
    \includegraphics[width=\linewidth]{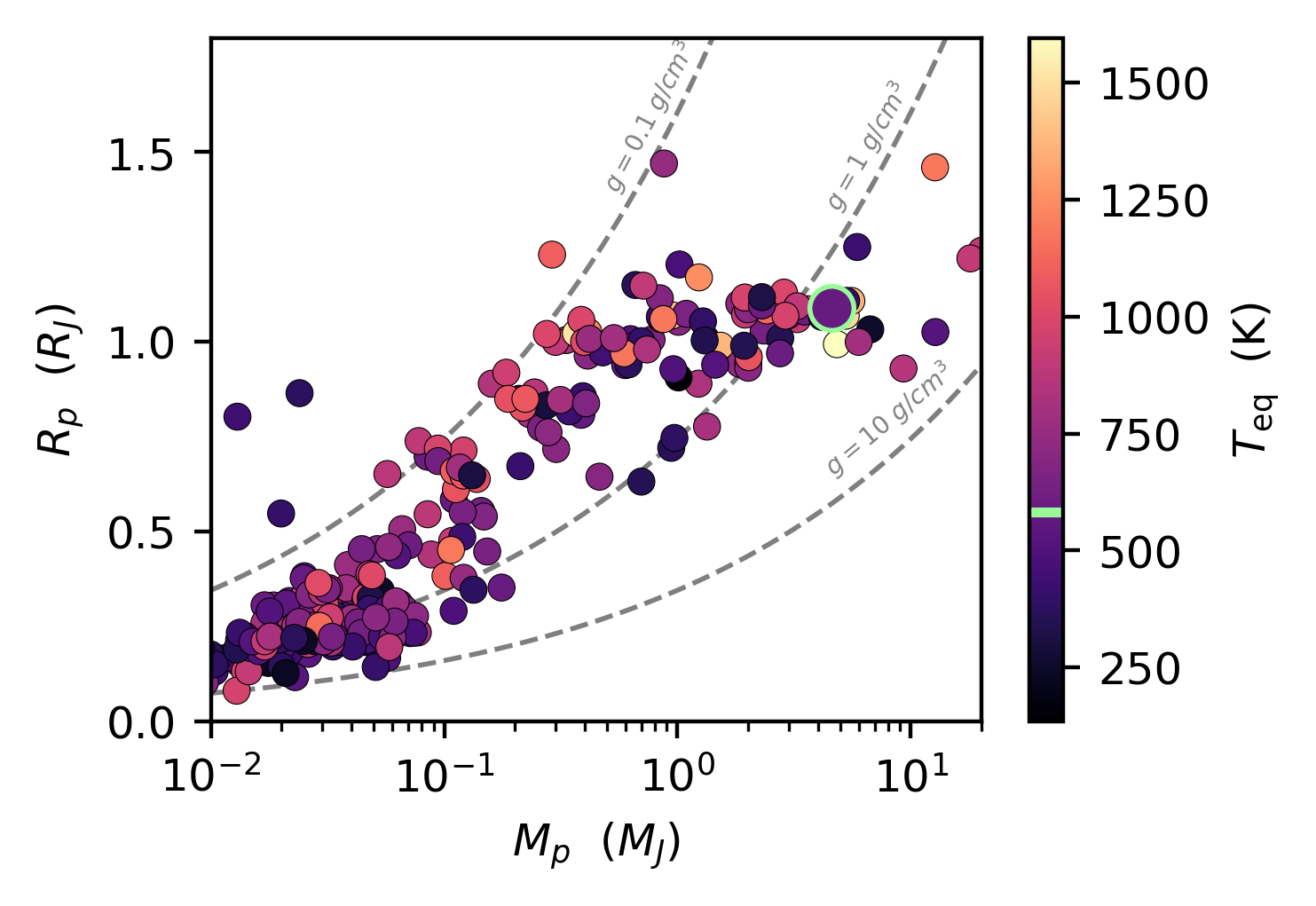}
    \caption{Mass vs radius diagram for confirmed planets with $P>10$ days. Circles are colored according to each planet's equilibrium temperature reported in the catalog.}
    \label{fig:MvsR}
\end{figure}

In terms of mass and radius, \plname\ is similar to Kepler-432 b \citep{kepler432}, Kepler-1704 b \citep{kepler1704}, TIC 241249530 b \citep{tic2412}, TIC 393818343 b \citep{tic3938} and TOI-2497 b \citep{toi2497}, however, the first 4 planets orbit in very eccentric orbits ($e>0.5$). TOI-2497 b has a similar mass, radius, and eccentricity, but has a period of $P=10.65$ days, making \plname\ quite unique in this regime of mass. This comparison is illustrated in Figure \ref{fig:MvsR}, putting \plname\ close to the upper limit in mass of current discoveries.

\begin{figure*}
    \centering
    \includegraphics[width=0.49\linewidth]{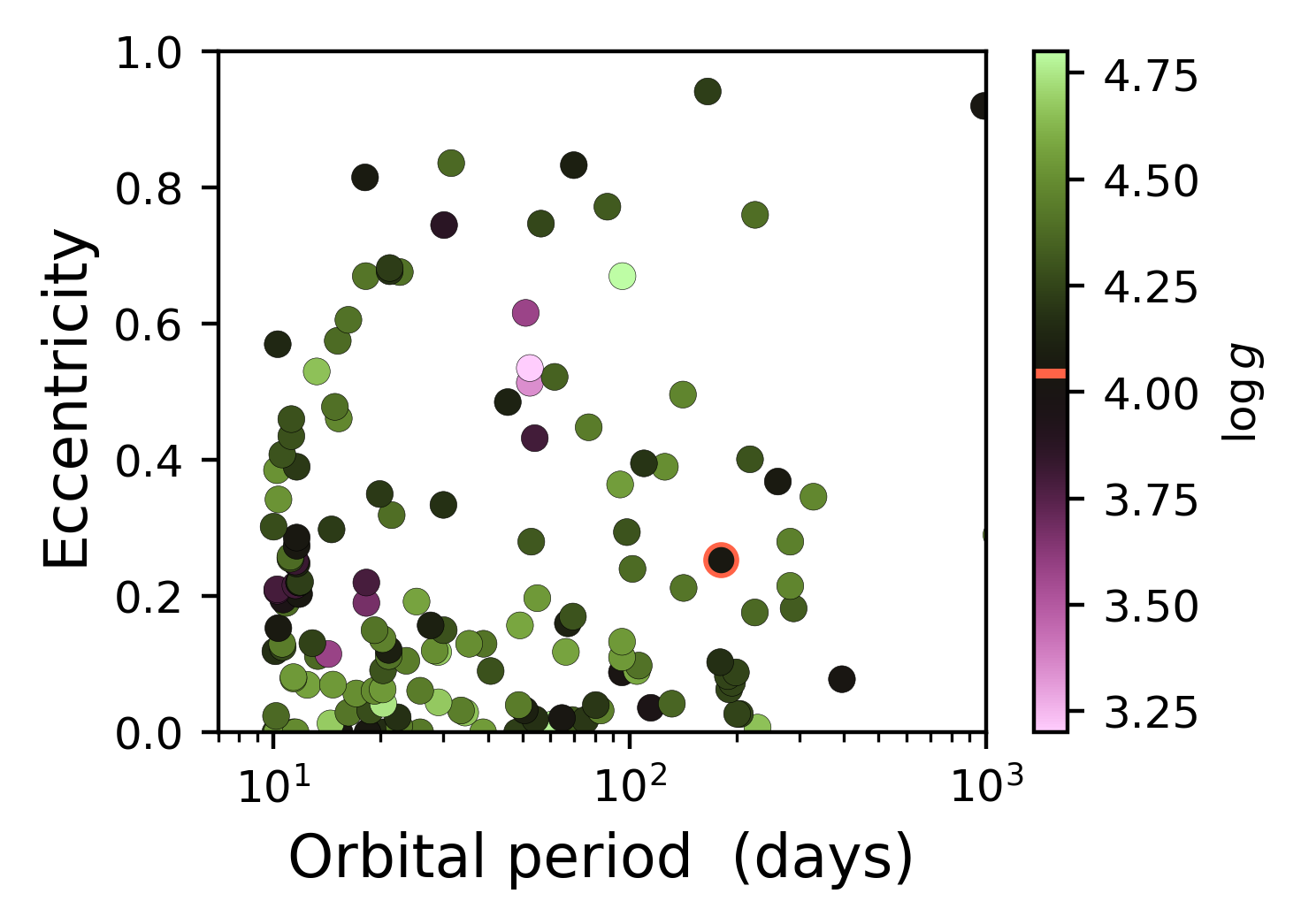}
    \includegraphics[width=0.49\linewidth]{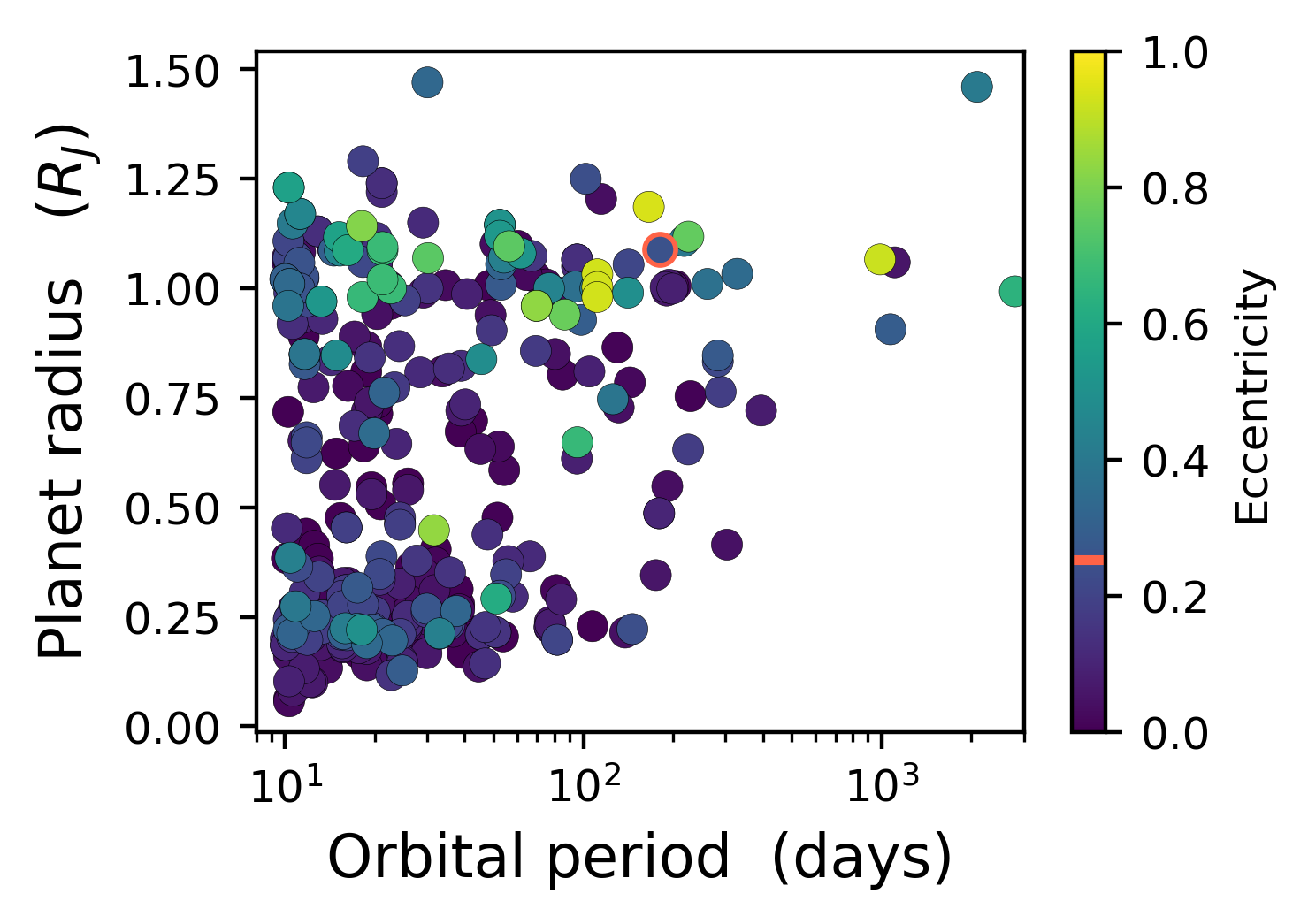}
    \caption{Planets with $P>10$ days and their orbital period vs eccentricity and planetary radii. The left panel shows the dependency of orbital period on eccentricity. $\log g$ is plotted in color scale. The right panel shows the period-radius diagram with \plname\ being one of the longer period giant planets discovered.}
    \label{fig:PvseR}
\end{figure*}

In the period-eccentricity diagram (left panel Figure \ref{fig:PvseR}), we observe that \plname\ appears to be in an unpopulated region. If we add mass into the comparison, we identify TOI-4465 b \citep{toi4465} as a quite similar planet, but orbiting a G dwarf star. The color bar traces $\log g$ as a proxy with stellar evolution stage, showing that the current sample of planets around evolved stars has eccentric orbits \citep{grunblatt2023}.

In the period–radius diagram (right panel Figure \ref{fig:PvseR}), we find that this planet is among the few warm Jupiters with an orbital period exceeding 100 days and a mild eccentricity, contrasting with some similar planets with much more eccentric orbits, as already stated in the previous paragraph.

\begin{figure*}
    \centering
    \includegraphics[width=0.49\linewidth]{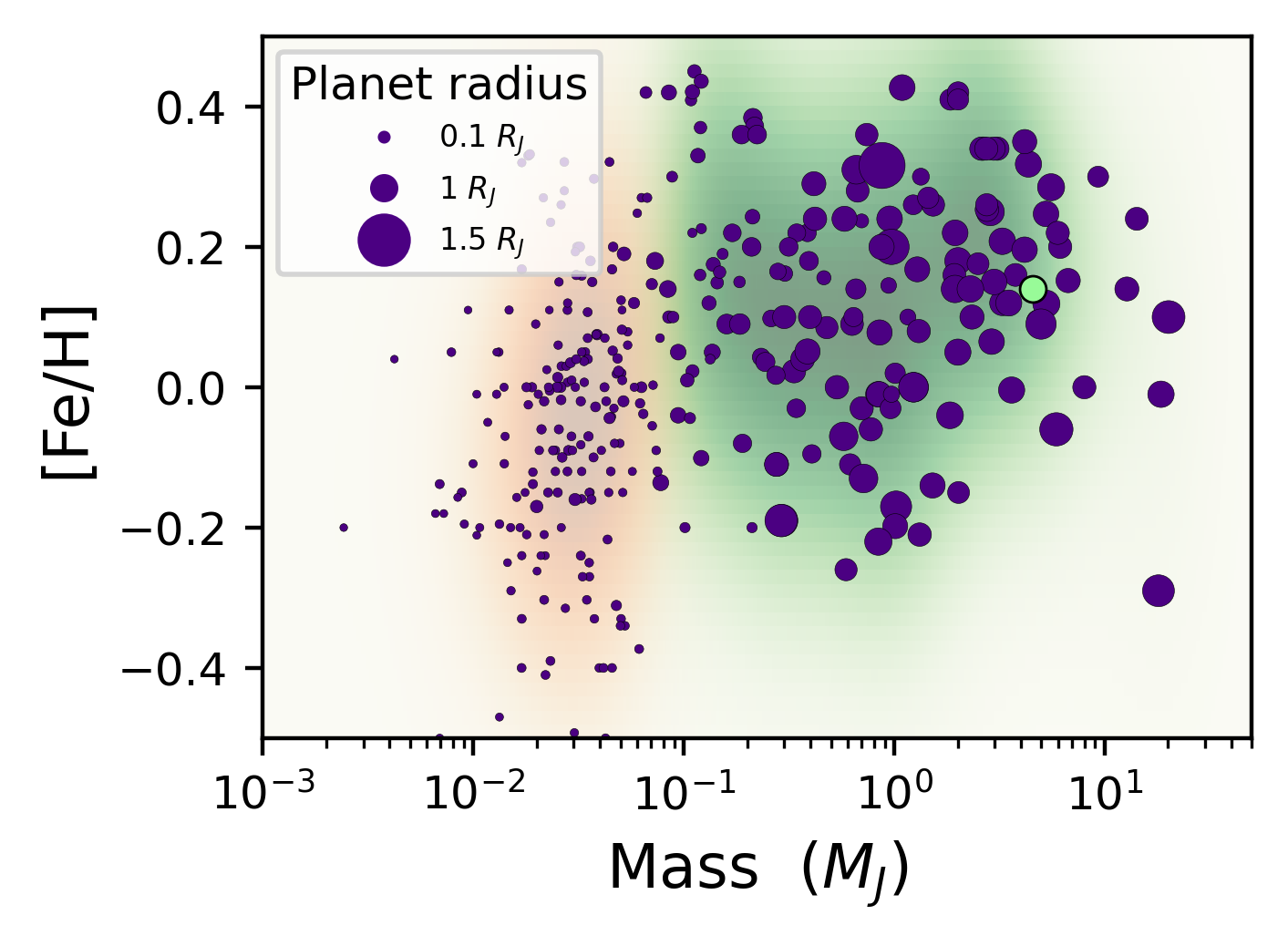}
    \includegraphics[width=0.49\linewidth]{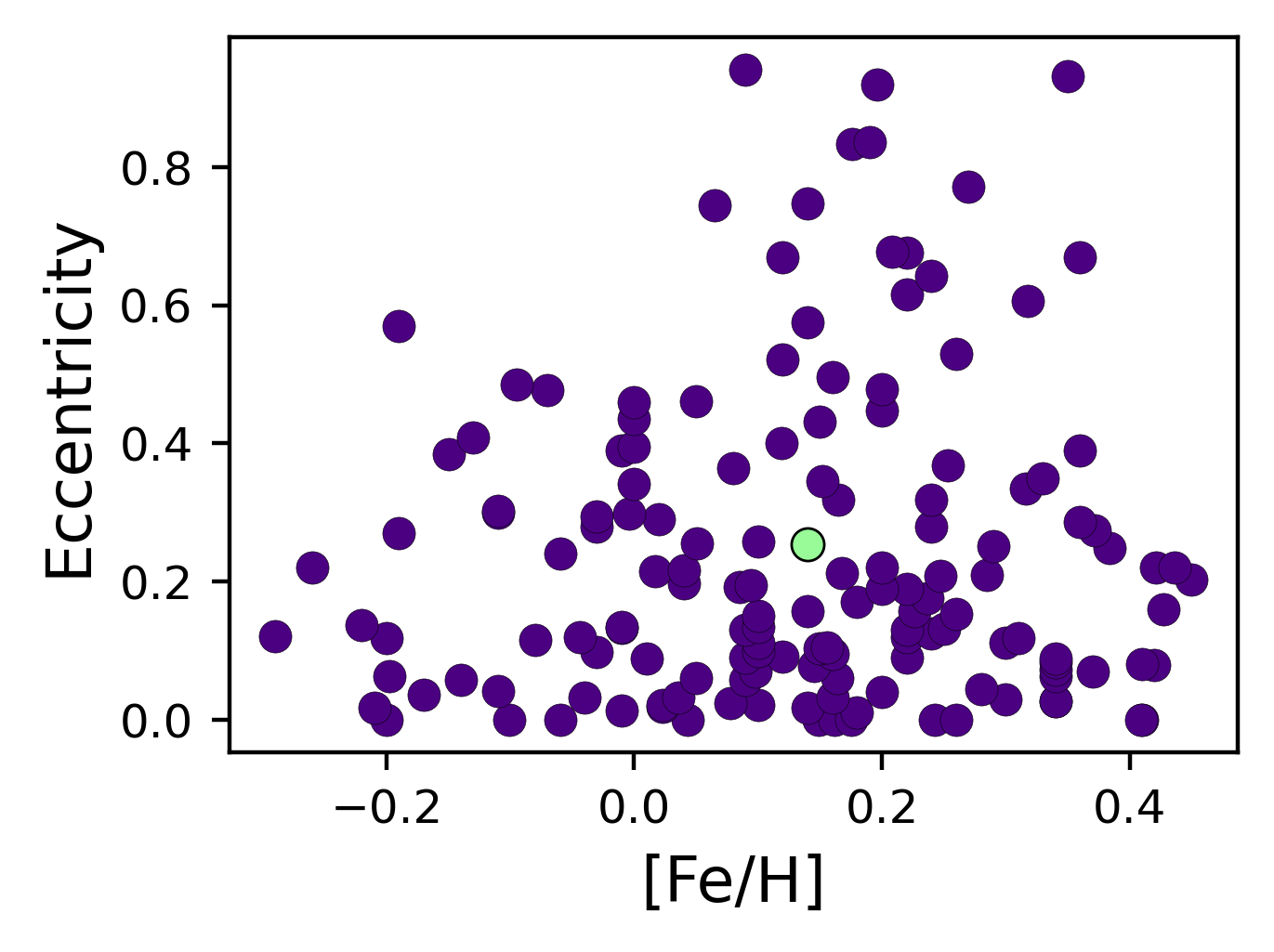}
    \caption{Left: Mass vs stellar metallicity diagram for planets with $P>10$ d. Kernel density estimation for planets with mass lower and higher than $0.1$ \mjup\ are plotted for reference. Right: Metallicity vs eccentricity diagram for planets with $P>10$ d and $M>0.1$ \mjup. \plname\ is shown in green in both panels.}
    \label{fig:Mvsmet}
\end{figure*}

Figure \ref{fig:Mvsmet} shows the mass and eccentricity dependence on stellar metallicity. The left panel shows a well-known correlation where giant ($M>0.1$ \mjup) planets are observed around metal-rich stars \citep{gonzalez1997,fischer2005,johnson2010}. \plname\ is no exception to this trend, being in an interesting place, as its mass around $4.5$ \mjup\ is very close to a proposed break, where more massive planets tend to be hosted by lower metallicity stars \citep{schlaufman2018, goda2019}. This separation would be related to different formation channels: the core-accretion mechanism \citep{coreac} in the lower mass regime and disk instability \citep{diskinsta, boss2024} for the most massive planets. The first mechanism is metallicity-dependent, and the second one requires a massive disk \citep{matsuo2007}. In contrast, \cite{adibekyan2019} doesn't find this separation in the giant planet population, but agrees on different formation mechanisms based on environmental conditions. The origin of giant planets with \mplanet $>4$ \mjup\ remains a debated topic \citep{matsukoba2023,nguyen2024}.

The right panel of Figure \ref{fig:Mvsmet} shows \plname\ in the metallicity-eccentricity space, where warm Jupiters exhibit a wider range of eccentricities at high metallicities \citep{dawson2013, alqasim2025, morgan2026}. \plname\ occupies a mid value in this range, and confirming the presence of unseen companions and/or a study of the obliquity would provide a better constraint on the migration scenario of this planet.

The interior modeling indicates a low level of heavy-element enrichment on this planet, similar to other massive giant planets, such as TIC4672985 b \citep{jones2024}. Additionally, the coreless model suggests that this planet may have formed through gravitational instability.

\subsection{Future observations}

The orbital obliquity can be measured through the Rossiter-McLaughlin (RM) effect \citep{rossiter,mclaughlin,rmeq}, which provides insight into the planet's migration scenarios \citep{petrovich2016}. In principle, an aligned orbit points to a disk migration scenario; meanwhile, a misaligned orbit hints at a previous interaction with a companion \citep{winn2010}. We estimate the amplitude of the RM effect to be $12.3\pm1$ m/s, which should be attainable with current state-of-the-art instruments from the southern hemisphere. Nevertheless, the 11.6-hour transit duration makes this study quite expensive and difficult to observe.

We also calculated the transmission spectroscopy metric \citep[TSM][]{tsm}, which quantifies a planet's potential for atmospheric studies. It's important to note that this metric is defined with a scale factor that accommodates planets up to 10 Earth masses. Therefore, we compute it using the scale factor from the last bin. The resulting TSM is 3.44, which falls in the lower range, making \plname\ not a good candidate for atmospheric studies, despite its relevance due to its cooler equilibrium temperature \citep{fortney2021}. This outcome can be attributed to the planet's low equilibrium temperature \teq\ along with its higher mass \mplanet\ and star radius \rstar . 

Finally, we calculated the emission spectroscopy metric \citep[ESM][]{tsm}, which is similar to TSM but focused on the expected signal-to-noise (SNR) ratio for thermal emission characterization using the James Webb Space Telescope (JWST). For \plname , we obtain an ESM value of 4.8, which is below the recommended SNR of 7.5 for this kind of study.

\section{Conclusions}
\label{sec:conclusions}

Starting from a single transit event and following up with ground-based photometric and spectroscopic observations, we report the discovery and characterization of \plname , a long-period warm Jupiter orbiting an evolved F-type star. With a period of \plperA\ days, mass of \mplA\ \mjup , radius of \rplA\ \rjup , eccentricity of \plecc, and equilibrium temperature of \teqA\ K, \plname\ joins the scarce population of warm Jupiters with periods longer than 100 days.

The Rossiter-McLaughlin amplitude of 12 m/s for this planet is suitable for observations with current facilities in terms of precision; however, the 11.6-hour transit duration makes it very challenging to observe in a single night. A low obliquity, pointing to an aligned orbit, would be quite significant, as it would support a disk-migration origin; meanwhile, most long-period warm Jupiters have extreme eccentricities, which are more related to a planet-planet scattering scenario.

\newpage
\begin{acknowledgements}

An independent study to confirm and characterize \plname\ was carried out and submitted simultaneously by Rodel et al. 2026, in prep.

This work has made use of data from the European Space Agency (ESA) mission
{\it Gaia} (\url{https://www.cosmos.esa.int/gaia}), processed by the {\it Gaia}
Data Processing and Analysis Consortium (DPAC,
\url{https://www.cosmos.esa.int/web/gaia/dpac/consortium}). Funding for the DPAC
has been provided by national institutions, in particular the institutions
participating in the {\it Gaia} Multilateral Agreement.

This paper makes use of data produced by the HATPI project(\url{https://hatpi.org}), located in Chile at Las Campanas Observatory of the Carnegie Institution for Science and operated by the Department of Astrophysical Sciences at Princeton University. External funding for HATPI has been provided by the Gordon and Betty Moore Foundation, the David and Lucile Packard Foundation, the Mount Cuba Astronomical Foundation, and the Agencia Nacional de Investigación y Desarrollo (ANID) of Chile through QUIMAL, Millennium and Fondecyt grants.

FR acknowledges the support from the Vicerrectoría de Investigación (VRI) at the Pontificia Universidad Católica de Chile. RB acknowledges support from FONDECYT Project 1241963 and from ANID – Millennium Science Initiative – ICN12\_009. JJ is grateful that the publication could be produced within the framework of institutional support for the development of the research organization of Masaryk University. LV acknowledges the support from ANID Fondecyt no. 1211162, Fondecyt no. 1251299, and BASAL FB210003.

\end{acknowledgements}

\newpage 

%

\bibliographystyle{bibtex/aa} 
\bibliography{biblio} 

\end{document}